\documentclass[twocolumn,aps,amsfonts]{revtex4}
\usepackage{epsfig}
\usepackage{bm}
\newcommand{\sfr}[2]{{\textstyle\frac{#1}{#2}}}
\newcommand{\reff}[1]{(\ref{#1})}

\begin{document}

\title{The modified Newtonian dynamics---MOND
\\
and its implications for new physics}

\author{Jacob D. Bekenstein}\thanks{\emph{Author's address:} Racah
Institute of Physics, Hebrew University of Jerusalem, Givat Ram,
Jerusalem 91904, Israel;\\ Email: bekenste@vms.huji.ac.il}

\begin{abstract} 
No more salient issue exists in contemporary astrophysics and cosmology than that of the elusive ``dark matter''.  For many years already Milgrom's paradigm of modified Newtonian dynamics (MOND) has provided an alternative way to interpret observations without appeal to invisible dark matter.  MOND had been successful in elucidating economically the dynamics of disk galaxies of all scales, while doing less well for clusters of galaxies; in its original form it could not address gravitational lensing or cosmology.  After reviewing some of the evidence in favor of MOND, I recollect the development of relativistic formulations for it to cope with the last deficiency.  I comment on recent work by various groups in confronting T$e$V$e$S, a relativistic embodiment of MOND, with observational data on gravitational lensing and cosmology. Throughout I ask what sort of physics can be responsible for the efficacy of MOND, and conclude with an appraisal of what theoretical developments are still needed to reach a full description of the world involving no unobserved matter.

\end{abstract}


\maketitle

\section*{\label{sec:intro}  1. Introduction} 
             
Newtonian gravity theory has served physics and technology faithfully for well over three centuries.   Nevertheless, it has long been known that it is only an approximation to a relativistic gravitation theory, usually identified with Einstein's 1915 general relativity (GR).  GR  has correctly predicted subtle effects in the dynamics of the solar system, in the celebrated Hulse-Taylor double pulsar, and has anticipated the existence of those exotic denizens of the universe,  black holes, now confirmed by myriad observations of galactic nuclei, compact galactic X-ray sources, etc.  The nonlinear aspects of gravitation in GR are crucial to these new settings: gravitation is more difficult than Newton supposed.  But up until a few decades ago astronomers took comfort in the belief that if one sidesteps situations with extremely strong gravitational fields, or with extremely rapid motions, then Newtonian theory  is  a good approximation to the truth.

But is that so?   Consider the situation in the realm of galaxies.  Speeds are low compared to light's, and gravitational fields are weak there.  So Newtonian physics \emph{should} describe motions extremely well.  Yet the accelerations of stars and gas clouds in the outskirts of spiral galaxies, and of galaxies swimming in the large clusters of galaxies,  well exceed Newtonian predictions made on the basis of the matter actually visible in these systems. Those accelerations are inferred  by combining Doppler measured velocities with geometric assumptions, e.g. that in a particular spiral galaxy stars move in circles on a plane.  The geometric assumptions can be checked, at least statistically, and the acceleration discrepancy is found to be real and generic.  Sometimes the problem is characterized thus: rotation or random velocities are much bigger than expected.  I stress acceleration because it is the quantity which directly measures the strength of the gravitational field.   In any case, what is wrong? 

It was realized by Ostriker and Peebles~\cite{OP} that massive halos would stabilize disk galaxies against the bar forming instabilities which, in Newtonian theory, are endemic to systems with little velocity dispersion.   Since about third of the spiral galaxies have no significant bar, this justified the hypothesis that disk galaxies (including spirals) are often embedded in massive but invisible halos.  Thus started the trend to resolve the acceleration discrepancy by postulating the existence of much dark matter (DM) in systems ranging from the very tenuous dwarf spheroidal galaxies with visible masses $\sim 10^7M_\odot$ to the great clusters of galaxies with observed masses in the $10^{14}M_\odot$ ballpark~\cite{DMU}, in brief in any system where an acceleration discrepancy exists.   The DM's role is to provide the missing gravitational pull to account for the excessive accelerations (an easy introduction to the DM paradigm is provided by Khalil and Mu\~ noz~\cite{KhM}).  But thirty years of astronomical exploration and laboratory experiments~\cite{expts} have yet to provide independent  evidence of DM's existence, e.g. $\gamma$ rays from its decay.  Is there an alternative to DM?   This review  focuses on the modified Newtonian dynamics scheme put forward by Milgrom~\cite{M1,M2,M3} as an alternative to the systematic appeal to DM.

\section*{2. Sharpening the problem}
 
\begin{figure}[ht]
\centering
\includegraphics[width=7cm]{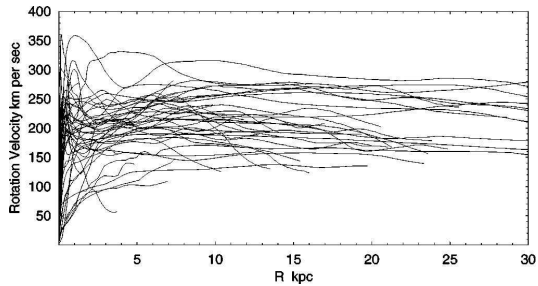}
  \caption{Collage of RCs of nearby spiral galaxies obtained by combining Doppler data from CO molecular lines for the central regions, optical lines for the disks, and HI 21 cm line for the outer (gas) disks.  The rotation velocity in units of km s$^{-1}$ is plotted vs galactocentric radius $R$ in kiloparsecs (kpc); 1 kpc $\approx$ 3000 light years.  It is seen that the RCs are flat to well beyond the edges of the optical disks ($\sim 10$ kpc).  Graph from Ref.~\onlinecite{SofRub}, reprinted with permission from the Annual Review of Astronomy and Astrophysics, Volume 39 (c)2001 by Annual Reviews www.annualreviews.org}
 \label{figure:fig1}
 \end{figure}

To pass judgment on DM and alternatives to it, it is well to take stock of the two overarching empirical facts in the phenomenology of disk galaxies.  First, ever since the work of Bosma as well as Rubin and coworkers~\cite{Rubin}, it has been clear that gas clouds in the disks of spiral galaxies, which serve as tracers of the gravitational potential, circle around each galaxy's center with a (linear) velocity which first rises as one moves out of the center, but hardly drops as the radius grows to well beyond the visible disk's edge. Yet a falloff of the velocity with radius was naively expected because, to judge from the light distribution in spiral galaxies, most of their visible mass is rather centrally concentrated.  Newtonian gravitation would thus predict that ``rotation curves'' (RC) should  drop as $r^{-1/2}$ outside the bright parts of these galaxies,  but this is not seen in over a hundred extended RCs measured with sufficient precision~\cite{SofRub,SMG,McGaughEdin}.  In fact in many spiral galaxies, particularly those possessing high surface brightness, the extended RC becomes flat away from the central parts (FIG.~\ref{figure:fig1}).

\begin{figure}[ht]
\centering
\includegraphics[width=6cm]{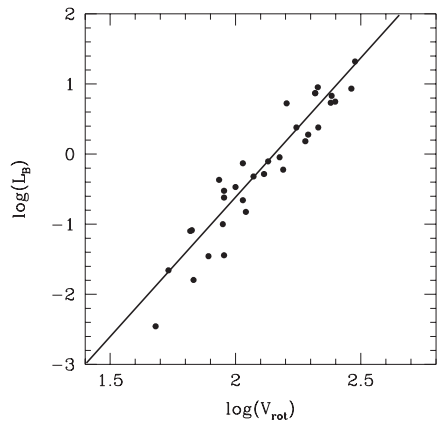}
  \caption{The TF correlation for a sample of galaxies as a log-log plot  of the blue band luminosity $L_{\rm B} $ in units of $10^{10}\, L_\odot$ vs. the asymptotic rotational velocity $V_{\rm rot}$  in km/s.    The straight line (mine) has slope exactly 4 and corresponds to MOND's prediction, Eq.~\reff{TFlaw}, for a reasonable mass (in units of ${\cal M}_\odot$)  to $L_{\rm B}$ ratio $\Upsilon_B=  3$. Graph reproduced from Ref.~\onlinecite{Sanders96} by permission of the American Astronomical Society.}
 \label{figure:fig2}
 \end{figure}

Second, as originally pointed out by Tully and Fisher (TF)~\cite{TF}, the rotation velocity $V_{\rm rot}$ and the blue band luminosity $L_{\rm B} $ of a galaxy  are simply correlated (FIG.~\ref{figure:fig2}).  In a more informative form~\cite{SMG},  the TF law states that for disk galaxies the luminosity in the near infrared band, $L_{K'}$  (a good tracer of stellar mass), is proportional to the fourth power of the rotation velocity in the flat part of the RC, with a universal proportionality constant.  

The flat extended RC's have provided the logic underpinning the DM paradigm as follows.  In the flat part of a RC, the rotation velocity is $r$ independent, so the centripetal acceleration goes as $1/r$.  Thus  in the plane of the galaxy the gravitational field must be decreasing as $1/r$.  According to Poisson's equation of Newtonian theory, such a gravitational field, if assumed spherically symmetric, must be sourced by a mass distribution with $1/r^2$ profile (isothermal sphere model).  Since the visible mass density (in stars and gas) in the inner disk drops much faster than this, it is consistent to assume that  the total mass distribution in the outer parts is quasispherical.  The conclusion is that each spiral galaxy must be immersed in a roundish DM halo with mass density profile tending, at large $r$, to $1/r^2$.  

But questions plague the halo hypothesis. The halo DM, though much searched  for, has never been detected directly~\cite{expts}.   Cosmogonic simulations of DM halo formation within Newtonian physics predict that such a halo would have a density profile behaving like $1/r$ for small $r$ and gradually steepening to $1/r^3$ asymptotically.  Only by fine tuning the halo's parameters is it feasible to match this kind of profile to the observed flat RCs as well as would the isothermal sphere (the need to fine tune parameters of a halo model in order to fit the shape of the galaxy's RC was already clear long ago~\cite{BahcCass}).  And the predicted $1/r$ ``cusp'' in the density profile is also observationally problematic~\cite{expts}.

The DM halo hypothesis is evidently an attempt to resolve the acceleration discrepancy within orthodox gravitation theory.   But in the coming to terms with this discrepancy, suspicion fell on Newtonian gravity already early in the game. Zwicky, who had exposed the acceleration discrepancy in clusters of galaxies~\cite{zwicky}, opined much later that the discrepancy may reflect a failure of conventional physics~\cite{Morpho}.  Concrete proposals for departures from the Newtonian inverse square law in various settings were put forward by a number of workers~\cite{Finzi}.  All of these still regarded gravitation as a linear interaction with the strength of the field proportional to its source's mass.  

As Milgrom realized~\cite{M1}, such a modification of the gravity law, if relevant on the scales of galaxies, is incompatible with the TF law.  For one it would imply that a mass $m_1$ at ${\bm r}_1$ generates at ${\bm r}_2$ the acceleration field $m_1 {\bm g}({\bm r}_2-{\bm r}_1)$, with ${\bm g}$ depending only on fundamental constants and on the relative positions of source and field points in accordance with the principle of homogeneity of space.  Then, obviously, for matter orbiting with speed $V_{\rm rot}$ at radius $R$ in the outskirts of a galaxy (with its source mass, $M$, concentrated at a fairly definite distance $R$), the acceleration, $V_{\rm rot}{}^2/R$ should be equal to  $M |{\bm g}(R)|$.  This would require not only $|{\bm g}(R)|\propto R^{-1}$, but also $M\propto V_{\rm rot}{}^2$, the last incompatible with the TF law.  Then again ${\bm g}({\bm r}_2-{\bm r}_1)$ with the correct dimensions, $L T^{-2} M^{-1}$, cannot be built by using only $G, {\bm r}_2-{\bm r}_1$ and the scale of length $\ell$ at which gravity departs from Newtonian form.  And inclusion of mass in the construction (presumably the source's mass)  would be against the spirit of linearity.  The bottom line is that linear gravity, even if non-Newtonian, is incompatible with the TF law.

\section*{3. The MOND paradigm}

\begin{figure}[ht]
\centering
\includegraphics[width=6cm]{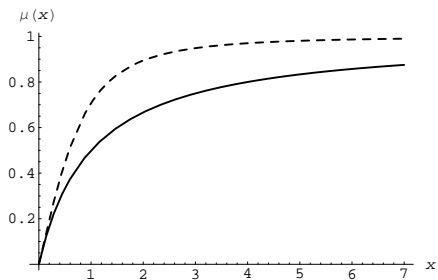}
  \caption{Two popular choices  for Milgrom's $\mu$ function: the ``simple'' function $\mu(x)=x/(1+x)$ (solid) and the  ``standard'' function $\mu(x)=x/\sqrt{1+x^2}$ (dotted).}
 \label{figure:fig3}
 \end{figure}

Milgrom~\cite{M1,M2,M3} proposed a novel paradigm which can be interpreted as reflecting non-Newtonian as well as nonlinear character of gravity already at the nonrelativistic level.  He called the scheme  Modified Newtonian Dynamics (MOND).
The essential part of it is the relation 
\begin{equation}
\mu(|{\bm a}|/a_0)\,{\bm a}=-\bm\nabla\Phi_N
\label{MOND}
\end{equation}
between the acceleration ${\bm a}$ of a particle and the ambient conventional Newtonian gravitational field $-\bm\nabla\Phi_N$. If the function $\mu$ were unity, this would be usual Newtonian dynamics.  Milgrom assumes that the positive smooth monotonic function $\mu$ approximately equals its argument when this is small compared to unity (deep MOND limit), but tends to unity when that argument is large compared to unity (FIG.~\ref{figure:fig3}).   The $a_0$ is a natural constant, approximately equal to $10^{-10}\,  {\rm m\, s}^{-2}$.  It is a fact that the centripetal accelerations of stars and gas clouds in the outskirts of spiral galaxies tend to be below $a_0$.

How does the MOND formula~\reff{MOND} help?  Consider again stars or gas clouds orbiting in the disk of a spiral galaxy (mass $M$) with speed $V(r)$ at radius $r$ from its center.  We must identify  $|{\bm a}|$ with the centripetal acceleration $V(r)^2/r$.  Sufficiently outside the main mass distribution we may estimate $|\bm\nabla\Phi_N|\approx GM/r^2$.  And at sufficiently large $r$, $|{\bm a}|$ will drop below $a_0$ and we shall be able to approximate $\mu(x)\approx x$.  Putting all this together gives $V(r)^4/r^2\approx  a_0 GM/r^2$ from which we may conclude two things.  First, $V(r)$ well outside the main mass distribution becomes independent of $r$, that is, the RC flattens at some value $V_f$.  Second, from the coefficients follows
\begin{equation}
M= (G a_0)^{-1}V_f{}^4.
\label{TFlaw}
\end{equation}
Introducing the ratio of mass to luminosity $L$, $\Upsilon$, we have the added prediction  $L=(G a_0\Upsilon )^{-1}V_f{}^4$.  Although $\Upsilon$ in the blue band varies somewhat with galaxy color, this last results explains the TF law, FIG.~\ref{figure:fig2}.    Theoretically $\Upsilon$ is much less variable in the near infrared (K') band~\cite{deJong}, which accounts for the extra sharpness of the TF relation in that band (see Fig.~2 in Ref.~\onlinecite{SMG}).

\begin{figure}[ht]
\centering
\includegraphics[width=6cm]{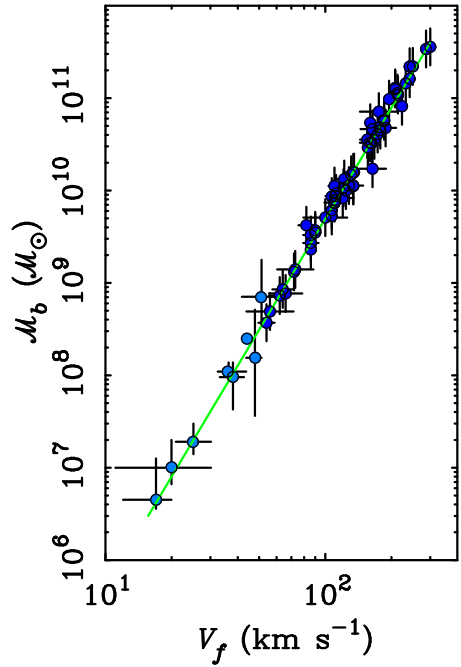}
  \caption{The baryonic TF correlation for galaxies spanning 6 orders of magnitude in mass. This is a log-log plot  of the total baryonic (stars plus gas) mass ${\cal M}_b$  in units of the solar mass ${\cal M}_\odot$ vs. the observed asymptotic rotational velocity $V_f$ in km/s.    For each of the 60 galaxies from Ref.~\onlinecite{SMG} (deep blue circles) the mass \emph{in stars} comes from a fit of the \emph{shape} of the RC with MOND, whereas for the eight dwarf spirals (light blue circles) the mass in stars (relatively small)  is inferred in Ref.~\onlinecite{McGaugh} directly from the luminosity.  The green line, with slope 4, is MOND's prediction, Eq.~\reff{TFlaw}.  Graph reproduced from Ref.~\onlinecite{McGaugh} by permission of the American Astronomical Society.}
 \label{figure:fig4}
 \end{figure}

The impressive agreement of Eq.~\reff{TFlaw} with observations is most clear from the baryonic  TF law~\cite{McGaugh}, FIG.~\ref{figure:fig4}.  In full agreement with MOND, the observed baryonic mass in spiral galaxies is accurately proportional to the fourth power of the asymptotic rotation velocity.   MOND's single formula thus unifies the two overarching facts of spiral galaxy phenomenology.

\begin{figure}[ht]
\centering
\includegraphics[width=8cm]{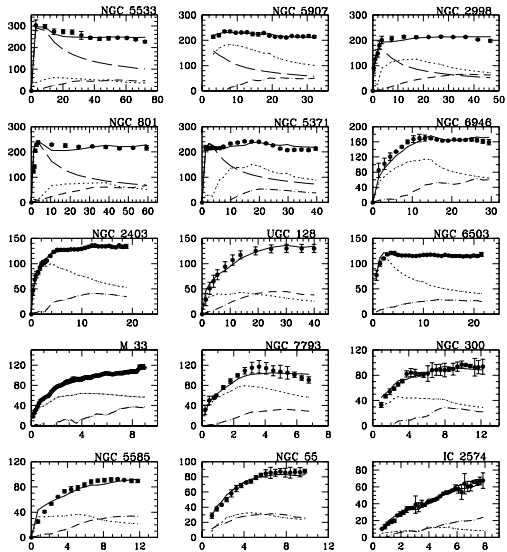}
  \caption{MOND fits to measured RCs of a sample of spiral galaxies  The galactocentric radius (horizontal axis) is in kpc and the rotation velocity (vertical)  in km/s. Each solid line is the MOND fit to the RC based on the distribution of light and neutral hydrogen.  The other curves are the Newtonian RCs contributed by the  gas (H and He) in the disk (short dashed), by the stars (dotted) and by the central bulge (if present, long dashed).   The free parameter of the fit is the mass of the stars in the disk (and the mass of the bulge if present).  Figure reproduced from Ref.~\onlinecite{SMG} with permission, from the Annual Review of Astronomy and Astrophysics, (c)2002 by Annual Reviews,  www.annualreviews.org.}
 \label{figure:fig5}
 \end{figure}
 
MOND is similarly successful in explaining the detailed \emph{shapes} of RCs.  
FIG.~\ref{figure:fig5} shows the measured RCs of fifteen galaxies together with the MOND fits based on surface photometry in the optical band and radio measurements.  The only parameter adjusted in these impressive fits is the disk's stellar mass, or equivalently, the stellar mass-to-luminosity ratio $\Upsilon$.  The trend of the so determined values of $\Upsilon$ with galaxy color jibes with that predicted  by stellar evolutionary models~\cite{deJong}; in this MOND does better than DM halo models~\cite{BKS}.  To make similarly successful fits without just ``putting DM where it is needed'', DM theorists must adjust two extra parameters apart from $\Upsilon$ in standardized halo models.  As stressed in a recent critical reappraisal~\cite{BKS},  those galaxies where one-parameter MOND fits do worse than halo fits often display complicating factors that could mitigate its less than striking performance.  In disk galaxies MOND is unquestionably more economical, and thus more falsifiable, than the DM paradigm.  

When applied to many galaxies, the MOND fits mentioned serve to determine $a_0$ with some accuracy; the value $a_0=1.2\times 10^{-10}\,  {\rm m\, s}^{-2}$ is often used~\cite{BBS}.  It is significant that  $a_0$ as deduced from RC fits agrees well with the value for $a_0$ obtained by comparing $(G a_0\Upsilon )^{-1}$ with the empirical coefficient in the TF law: the roles of $a_0$ are different in the two subjects, and they are tied together only in MOND.  Milgrom~\cite{M2002} notes a few additional conceptually distinct roles played by $a_0$ in extragalactic astrophysics. An early example is that $a_0$ sets a special scale of mass surface density $\Sigma_m=a_0/G$, and wherever in a system the actual surface mass density drops below $\Sigma_m$, Newtonian gravitational behavior gives way to MOND dynamics~\cite{M2}. (In galaxies like ours this occurs some way out in the disk, and that is why the RC may exhibit a brief drop before becoming flat asymptotically).  On this basis Milgrom predicted that were disk galaxies with surface mass density everywhere below $\Sigma_m$ to exist, they should show especially large acceleration discrepancies.   A population of such galaxies became known in the late 1980's~\cite{BBS}, and all facets of Milgrom's prediction were subsequently confirmed~\cite{McGdB}.  MOND obviously predicts that in  these so called low surface brightness galaxies, the shapes of the RCs, which tend to be \emph{rising} throughout the visible disk, should be independent of the precise way $\mu(x)$ switches from $x$ to unity for $x\sim 1$.

Constraints of space forbid me from delving into other MOND successes, e.g. for elliptical and dwarf spheroidal galaxies. But excellent reviews of these and other phenomenological aspects of MOND exist~\cite{SMG,McGaugh2,Scarpa}.  It is well, however, to stress one prominent empirical difficulty faced by the MOND formula~\reff{MOND}.  

Clusters of galaxies exist which comprise hundreds of galaxies of various kinds moving in their joint gravitational field with velocities of up to $10^3\, {\rm km\, s}^{-1}$.  The Newtonian virial theorem (gravitational potential energy of a quiescent system equals minus twice its kinetic energy) can be used to estimate a cluster's mass since the kinetic term is proportional to the total mass while the potential one goes like the square of mass. In this way it is determined that the total gravitating mass in a typical cluster is 5-10 times the mass actually seen in galaxies as well as in hot X-ray emitting gas, an oft constituent of clusters.  A similar story is told by analyses of the hydrostatics of the hot gas in light of its measured temperature.  Finally, the gravitational lensing by several clusters has led to estimates of their masses commensurate with the mentioned diagnosis.

When the MOND formula, or even better, a MOND analog to the virial theorem~\cite{M94b}, is used to estimate the masses of clusters, one finds an improvement but not a full resolution to the acceleration discrepancy. Clusters still seem to contain a factor of two more matter than actually observed in all known forms~\cite{SMG}.  The optimist will stress that MOND has alleviated the discrepancy without even once overcorrecting for it; the pessimist~\cite{Silk} will view this finding as damaging MOND's credibility.  But one should keep in mind that clusters may contain much invisible matter of rather prosaic nature, either baryons in a form which is hard to detect optically, or massive neutrinos~\cite{S03,Pointecouteau}.  However, the option that clusters contain non-baryonic cold DM between the galaxies, while logically possible, seems hardly justifiable in view of MOND's overall philosophy.

\section*{ 4. What is behind the MOND formula?}
\label{behind}

What is the physical basis of MOND's success in unifying a lot of extragalactic data?  There are certain things the MOND formula cannot be.  First, it is unlikely to be just a recipe for the way DM is distributed spatially in astronomical objects, as sometimes proposed~\cite{Turner}. Such proposals strive to explain how the scale $a_0$ enters into spiral galaxy properties through some regularity in the cosmogony of DM halos.  But strong arguments exists against an intrinsic scale of the halos~\cite{SB}.  And it is hard to see how such an explanation could give an account of the multiple roles of $a_0$ in different systems~\cite{M2002}. 

The MOND formula, conceived as \emph{exact},  cannot be a generic modification of the inertia aspect of Newton's second law $m{\bm a}={\bm f}$ (here ${\bm f}$ is the sum of \emph{all} forces acting on the particle including the Newtonian gravity force), an alternative proposed originally by Milgrom~\cite{M2} (see a latter elaboration in Ref.~\onlinecite{M94}).  To see why consider a binary system with unequal masses $m_1$ and $m_2$ evolving under mutual gravity alone.  The time derivative of $m_1{\bm v}_1+m_2{\bm v}_2$, as calculated from Eq.~\reff{MOND}, does not vanish in general if at least one of the accelerations is comparable to or smaller than $a_0$, since the $\mu$'s will generally not be equal.  So in the proposed interpretation, the MOND formula does not conserve momentum.  By the same token it does not conserve angular momentum, nor energy.  

Milgrom sidesteps the problem by stipulating that the MOND formula is only valid for test particles moving on a given background, e.g., stars moving in the collective gravitational field of a galaxy.    We may inquire more generally, does the MOND formula, or some closely related one, represent  a modification of inertia in test particle motion?   To comply with the conservation laws we likely want the formula to arise from a Lagrangian.  The kinetic part of the Lagrangian should give us the $\mu(|{\bm a}|/a_0)\,{\bm a}$ part with whatever ``corrections'' are required.  However, it is a theorem that no MOND-like dynamics exists that simultaneously has a Newtonian limit for $a_0\to 0$ (all accelerations are large), is Galilei invariant, and is derivable from a \emph{local} action~\cite{M94}.  (``Local'' means that the relevant Lagrangian can be written as a single integral over volume.)   This prohibition is even more stringent from a relativistic standpoint~\cite{Domokos}.  Accordingly  Milgrom introduced a \emph{nonlocal} action, i.e. one which is a functional of complete orbits, but cannot be reduced to an integral over a Lagrangian~\cite{M94}.   This approach does not  quite reproduce formula~\reff{MOND} generically, but that formula is recovered for the special case of circular orbits.  

Of course circular stellar orbits \emph{a la} MOND is really all that one needs to analyze data on spiral galaxies.  However, it is known that elliptical galaxies comprise highly radial stellar orbits too, yet elliptical galaxies also seem to be well described by MOND~\cite{MS}, and so the nonlocal modified inertia approach might be found wanting here.  On the plus side one may mention the Pioneer anomaly~\cite{Anderson}.  The Pioneer 10 and Pioneer 11 spacecraft, as they travelled almost radially in the outer parts of the solar system, were found to be subject to an anomalous sunward acceleration of order $8\times 10^{-10} {\rm m\, s}^{-2}$ (not far from $a_0$) which does not fall off measurably with distance from the sun. Were such an acceleration to reflect a generic gravitational field, it would have affected the outer planet ephemerides to an intolerable extent~\cite{Iorio}.   (An ephemerides is the calculated position of a celestial body as a function of time, past or future.)    By contrast, Milgrom's  nonlocal theory  does predict different modifications of Newtonian dynamics for radial (Pioneer) and circular (planetary) orbits, though details have yet to be worked out~\cite{M05}.  

Milgrom has also speculated~\cite{M99,M05} that modified inertia may have its origin in an effective interaction of bodies with the vacuum.  This is motivated by the well known fact that an object swimming through a fluid has its inertial mass increased by interacting with fluid degrees of freedom.  Of course, by local Lorentz invariance inertial motion through the vacuum is possible as usual.  However,  accelerated motion is distinguished from inertial motion (even before we come to inertia) by the  perception that the accelerated system is immersed in a thermal bath whose temperature is proportional to the acceleration (Unruh radiation~\cite{Unruh}).  Accordingly, Milgrom suggests that accelerated motion with respect to the vacuum may shape the inertial characteristics of objects in a way compatible with MOND.  An alternative speculation~\cite{M99,M05} is that MOND style  modified inertia and its particular scale $a_0$ may be generated by a symmetry in a higher dimensional spacetime, just as we think of ordinary (Newtonian or Einstenian) inertia as reflecting Lorentz symmetry in Minkowski spacetime.  In one implementation
of this program, with the extremely symmetric deSitter spacetime playing the role of the embedding spacetime, a relation between the cosmological constant and $a_0$ surfaces, which is actually approximately satisfied if the observed acceleration of the  Hubble expansion is interpreted as reflecting a nonzero cosmological constant.

\section*{5. MOND as modified gravity}

What if, instead of the modified inertia interpretation, we follow Milgrom's alternative proposal~\cite{M1} to regard MOND as a modification of Newtonian gravity?  This entails a change in the r.h.s. of $m{\bm a}={\bm f}$, with the new gravitational force possibly depending nonlinearly on its sources.   Can such implementation be protected from violation of the conservation laws?  

An affirmative answer to this query is readily available by starting from a modification of the Lagrangian of Newtonian gravity with Milgrom's $a_0$ playing the role of characteristic scale, to wit
\begin{equation}
L = -\int\Big[{a_0{}^2\over 8\pi G} F\Big({|\bm\nabla\Phi|^2\over a_0{}^2}\Big) + \rho\Phi\Big]d^3x.
\label{lagrangian}
\end{equation}
Here $F$ is some positive function, $\rho$ is the matter's mass density and $\Phi$, the field in the theory, is to be identified with the gravitational potential that drives motion, i.e. ${\bm a}=-\bm\nabla\Phi$.  The Lagrangian's kinetic part is the most general one that depends only on first derivatives of $\Phi$, and is consistent with the isotropy of space (only $|\bm\nabla\Phi|$ appears).  This Aquadratic Lagrangian theory (AQUAL)~\cite{BM} reduces to Newton's in the limit $F(X)\rightarrow X$, when it readily leads to Poisson's equation for $\Phi$.  More generically it yields the equation
\begin{eqnarray}
\bm\nabla\cdot[\mu(|\bm\nabla\Phi|/a_0)\bm\nabla\Phi]&=&4\pi G\rho,
\label{AQUAL}
\\
\mu(\surd X) &\equiv& F'(X).
\label{F}
\end{eqnarray}
One identifies the $\mu$ function here with MOND's $\mu$.

Comparison of the AQUAL equation~\reff{AQUAL} with Poisson's for the same $\rho$ shows that
\begin{equation}
\mu(|\bm\nabla\Phi|/a_0)\bm\nabla\Phi=\bm\nabla\Phi_N+\bm\nabla\times {\bm  h},
\label{integral}
\end{equation}
where ${\bm h}$ is some calculable vector field.    Now in highly symmetric situations  (spherical, cylindrical or plane symmetry), use of Gauss' theorem with the AQUAL equation shows that $\bm\nabla\times {\bm  h}$ has to vanish.  In light of the relation ${\bm a}=-\bm\nabla{\Phi}$, AQUAL then reproduces the MOND formula~\reff{MOND} exactly.  However, the curl term does not usually vanish in situations with lower symmetry (Milgrom and Brada exhibit a curious exception~\cite{Brada}).  We conclude that the MOND formula is generically an approximation to the solution of the AQUAL equation.  

How good an approximation?    Work by Milgrom and Brada~\cite{Mnumer,Brada} has shown that often the curl is just a 5-10\% correction.  This explains how the MOND formula manages to be so successful while not constituting by itself a physically consistent theory.  Now almost all MOND fits to observed RCs are calculated using that formula exclusively.  It should thus be clear that modest departures of empirical RCs from MOND's predictions do not constitute evidence against the paradigm.  The era of RC modeling with the full AQUAL Eq.~\reff{AQUAL} is still in its early stages~\cite{Ciotti}. 

I mentioned in Sec.~4 that the MOND formula is inconsistent with the conservation laws.  The AQUAl theory is free of this problem.  This needs no special proof (although explicit proofs exist~\cite{BM}): a theory based on a Lagrangian which does \emph{not} depend \emph{explicitly} on coordinates, directions and time (all true for AQUAL's Lagrangian) automatically conserves momentum, angular momentum and energy.  At the level of the first integral (\ref{integral}) of the AQUAL equation,  it is the curl term which protects AQUAL gravity from violating the conservation laws. 

AQUAL consistently embodies the weak equivalence principle (all bodies, regardless of inner structure, and starting from the same initial conditions, follow the same path in a gravitational field).  As remarked by Milgrom~\cite{M1}, the MOND formula is ambiguous in this respect.  In a star accelerations of individual ions greatly exceed $a_0$ in magnitude.  Thus for each ion  $\mu$ should be very close to unity, and we would guess that $\mu\approx 1$ for the star as a whole. Accordingly, if regarded as a collection of ions, the star would follow a Newtonian orbit in the galaxy's field.  This conclusion is entirely at variance with the evidence from the flat RCs, or from the dictate of the MOND formula when applied to the star regarded as structureless.  AQUAL predicts that the center of mass (CM) of a collection of particles, no matter what the internal gravitational field, will follow the orbit determined by $\bm a=-\bm\nabla\Phi$ with $\Phi$ the external potential coming from the AQUAL equation~\cite{BM}.  As mentioned, this acceleration will approximate that predicted by the MOND equation.  AQUAL does indeed supply a physical basis for MOND.

A system with sub-$a_0$ internal gravitational field immersed in an external field $g_e>a_0$ is predicted by AQUAL to exhibit quasi-Newtonian internal dynamics~\cite{BM}.  Originally Milgrom conjectured this ``external field effect'' to explain why no acceleration discrepancy is evident in the loose open star clusters residing in the Milky Way~\cite{M1}: the galaxy's field, still stronger than $a_0$ in its inner parts, suppresses the MOND behavior expected in light of the weak internal fields.   By contrast, if the said system  is immersed in an external field $g_e<a_0$, which is, however, \emph{stronger} than its internal fields, the internal dynamics is predicted to be quasi-Newtonian, but with the effective gravitational constant $G/\mu(g_e/a_0)$~\cite{BM}.

A useful product of AQUAL is a formula for the force between two masses in  motion about each other, e.g. a binary galaxy.  The MOND formula cannot be used directly here because, as mentioned in Sec.~4, it would suggest that the binary's center of mass should accelerate (nonconservation of momentum).  Instead, $\Phi$ has to be computed from the AQUAL equation, and then $\rho\bm\nabla\Phi$ has to be integrated over one galaxy's volume to obtain the force on it, as effectively done by Milgrom~\cite{Mnumer}.  For point masses $m_1$ and $m_2$ separated by distance $r$, the force is $(Gm_1m_2/r^2)f(m_2/m_1,r/r_t)$ with $r_t\equiv [G(m_1+m_2)/a_0]^{1/2}$; Milgrom gives the graph of $f$.  In the Newtonian limit ($r\ll r_t$), $f\approx 1$.   In the deep MOND limit ($r\gg r_t$), Milgrom~\cite{M94b} shows analytically that the force is $(2/3)(Ga_0)^{1/2}r^{-1}[(m_1+m_2)^{3/2}-m_1^{3/2}-m_2^{3/2}]$.  This is most elegantly computed by exploiting the discovery that in that limit the AQUAL equation is conformally invariant, an approach that also yields the exact virial relation in the deep MOND regime~\cite{M97}.

AQUAL permits a study of galaxy disc stability in the MOND paradigm.  I mentioned in Sec.~I that Newtonian disk instabilities motivated the idea that DM halos surround disk galaxies.  In the MOND paradigm there is not DM, so how do a fraction of the spiral galaxies in the sky avoid forming bars?   Using a mixture of MOND formula and AQUAL equation arguments, Milgrom~\cite{M89} showed  analytically that MOND enhances local stability of disks against perturbations as compared to the Newtonian situation.  Brada and Milgrom~\cite{BradaM} combined an $N$-body code with a numerical solver for the AQUAL equation to verify this and to demonstrate that MOND enhances global stability.  Both works conclude that the degree of stabilization saturates in the deep MOND regime; this means no disk in MOND is absolutely stable.   Christodoulou~\cite{Chr} and Griv and Zhytnikov~\cite{Griv} reach similar conclusions with $N$-body simulations based on just the MOND formula.    Recently Tiret and Combes~\cite{Tiret} studied \emph{evolution} of bars in spiral disks using $N$-body simulations built on the AQUAL equation.  They find that bars actually form more rapidly in MOND than in DM halo Newtonian models, but that MOND bars then weaken as compared to those in Newtonian models.  The long term distribution of bar strengths predicted by MOND is in better agreement with the observed distribution than that obtained from Newtonian halo models.

AQUAL is also important for studies of two-body relaxation in stellar systems.  Two-body relaxation is the process whereby distant gravitational encounters of pairs of stars in a galaxy or a cluster redistribute the system's energy and help it to approach some sort of equilibrium.  In Newtonian theory two-body relaxation is excruciatingly slow.   But as Binney and Ciotti show~\cite{BinneyCiotti}, two-body relaxation in the framework of AQUAL proceeds  faster.  In fact, if the system is deep in the MOND regime, the speedup factor is the square of the acceleration discrepancy.  

A related effect is dynamical friction, whereby a massive object moving through a collection of lighter bodies, e.g. a globular star cluster coursing among the stars of its mother galaxy,  loses energy by virtue of its kicking the background objects  gravitationally as it sweeps by them.  As a result the heavy object settles in the gravitational well confining it.  In AQUAL for systems deep in the MOND regime, dynamical friction is speeded up over the Newtonian value by one power of the acceleration discrepancy~\cite{BinneyCiotti}.

Both of the above phenomena open a new window onto the MOND-Newtonian gravity dichotomy.  \emph{Suppose} AQUAL is the correct nonrelativistic description of gravity.  In regard to RCs of spiral galaxies, Newtonian theory with DM could be ``saved'' by imagining the DM in each galaxy to be distributed in just such a way that the resulting total $\Phi_N$ is just the $\Phi$ generated in AQUAL theory by the baryonic (visible) matter.  Admittedly, the required DM distribution might be peculiar (negative density somewhere~\cite{M86}, or incompatible with cosmogonic simulations).  But leaving these options aside, observational uncertainties might make it difficult to tell  the two theories apart, as recent fits of RCs of nearby galaxies based on high quality data testify~\cite{Salucci}.  Yet consideration of the mentioned dissipative effects might remove the effective degeneracy between the two models of a galaxy offered by the two theories.

In view of the generic prediction of Newtonian cosmogonic simulations that the DM distribution in a galaxy rises to a cusp at the center (see Sec.~I),  Newtonian dynamical friction should have caused the globular clusters of many a dwarf spheroidal galaxy to already have merged at its center and there lost their integrity~\cite{Goerdt}.  While globular clusters are indeed absent from several such dwarf companions of the Milky Way, e.g. Draco, Sextans and Carina~\cite{Mateo}, the Fornax dwarf spheroidal has five globular clusters near, but \emph{not at}, its center.  Goerdt, Moore, et al.~\cite{Goerdt}  regard this as showing that the DM near the core of Fornax has a flat density profile, in contradiction to the DM simulations.  In fact, there is other growing evidence negating DM density cusps~\cite{Evans,expts}.  Now, some of the dwarf spheroidal satellites of the Galaxy including Fornax are in the deep MOND regime.   Following Binney and Ciotti~\cite{BinneyCiotti}, S\'anchez-Salcedo, Reyes-Iturbide and Hernandez~\cite{Sanchez}  point out that the mentioned speedup of settling by dynamical friction in AQUAL  should long ago have caused the globular clusters in Fornax to amalgamate at its center.  AQUAL thus has a problem explaining the observations of Fornax; however, as we see the DM alternative has its problems too.  

Can MOND give a picture of the late stages of galaxy formation?   A step in this direction was taken by Nipoti, Londrillo and Ciotti~\cite{Nipoti} on the basis their earlier work with $N$-body codes incorporating the full AQUAL equation~\cite{Ciotti}.  They simulate the dissipationless collapse of a cloud of particles which ends up in the MOND or deep MOND regimes. They note that phase mixing (the disappearance of  structure in the initial conditions due to commingling of particle orbits) proceeds slower than in the counterpart  Newtonian simulations, and that the final velocity distribution is anisotropic and favors radial orbits.  Current thought in cosmogony is that elliptical and spheroidal galaxies are the outcome of anisotropic collapse \emph{after} star formation is complete.  Nipoti et al. note that the end-products of some of their simulations resemble, in their surface density and velocity dispersion profiles, the dwarf-elliptical and dwarf-spheroidal galaxies in the sky.  But other final states have trouble fitting in all the accepted empirical correlations between luminosity, radius and velocity dispersion of elliptical galaxies.

\section*{6. Can MOND be made relativistic?}

Why is a relativistic version of MOND important?   Although most astronomical systems with significant acceleration discrepancy  are nonrelativistic in the extreme, there are two important exceptions.   Cosmology, which is important in itself and as a framework for the process of galaxy formation, is widely regarded as requiring the presence of much DM on the largest scales, and cosmology is universally believed to require relativistic treatment.  Gravitational lensing by  galaxies and clusters of galaxies has become a salient astronomical tool in the last two decades, and, since it discloses acceleration discrepancies, it is commonly regarded as supporting the need for DM.  Since light propagates with speed $c$, nonrelativistic MOND cannot even begin to formulate the problem of gravitational lensing.

The above cases drove the relativistic formulation of AQUAL~\cite{BM}, which I shall refer to as RAQUAL.  RAQUAL retains Einstein's equations of GR as a tool for deriving the spacetime metric $g_{\alpha\beta}$ from the energy-momentum tensor of the matter (one of whose components represents matter density).  RAQUAL further stipulates that matter and radiation play out their dynamics, not in the arena of $g_{\alpha\beta}$, but in that of $\tilde g_{\alpha\beta}=e^{2\psi/c^2} g_{\alpha\beta}$, where $\psi$ denotes a scalar field with dimensions of squared velocity. Since all matter is treated alike in this respect, the exquisitely tested weak equivalence principle is safeguarded, but since the metric governing gravitational field dynamics is different from the one experienced by matter, the strong equivalence principle fails.  This is essential, for otherwise we should end up back with GR as the relativistic gravity theory, and GR  in the nonrelativistic limit leads to Newtonian theory, not to MOND.

How is $\psi$ determined in RAQUAL?  One takes the Lagrangian for $\psi$ to be the covariant version of that for $\Phi$ in AQUAL, Eq.~\reff{lagrangian}.  This means replacing $|\bm\nabla\Phi|^2\mapsto g^{\alpha\beta} \partial_\alpha\psi\,\partial_\beta\psi$  as well as $d^3x\mapsto (-g)^{1/2}\,d^3x$, where $g$ is the determinant of the matrix of components of $g_{\alpha\beta}$.  But in RAQUAL one does not include a term in lieu of $\rho\Phi$; the coupling between $\psi$ and matter, which is to provide the source of the equation for $\psi$ in accordance with its Lagrangian, is automatically generated by the $e^{\psi/c^2}$ factor in the metric $\tilde g_{\alpha\beta}$ with which we built the \emph{matter}'s Lagrangian.   For time independent situations with nonrelativistic matter sources,  the $\psi$  equation  reduces to Eq.~\reff{AQUAL} with $\Phi\mapsto \psi$.  However, it would be a mistake to conclude that in RAQUAL $\psi$ is the gravitational potential.

In GR for weak gravity and nonrelativistic motion, the potential in which matter moves, $\Phi_N$,  is related to the metric  by $\Phi_N=-\sfr12(g_{tt}+c^2)$ ($t$ is the time coordinate). Because,  in this linearized theory,  the most relevant Einstein equation reduces to Poisson's, $\Phi_N$ coincides with the usual Newtonian potential.    In RAQUAL $g_{\alpha\beta}$ still satisfies Einstein-like equations, so the above relation applies as well. But the stipulation that matter's dynamics go forward in the metric  $\tilde g_{\alpha\beta}$ determines the nonrelativistic potential in which matter moves, now called $\Phi$, to be  $\Phi=-\sfr12(\tilde g_{tt}+c^2)$.  Writing in linearized theory  $\tilde g_{tt}=(1+2\psi/c^2)\,g_{tt}$, we have $\Phi=\Phi_N-g_{tt}\psi/c^2$.  Now to first approximation $g_{tt}=-c^2$; thus to leading order 
\begin{equation}
\Phi=\Phi_N+\psi.
\label{sum}
\end{equation}
Consequently, in RAQUAL the gravitational potential traced  by the matter's motions is the sum of the Newtonian potential and the AQUAL scalar field, both generated by the same baryonic matter.  The $\psi$ does the job of the DM's potential in conventional approaches, but its source is ordinary matter.

Let us compare RAQUAL with the successful AQUAL.  We take the $\mu$ in Eq.~\reff{AQUAL} with $\Phi\mapsto \psi$ to be approximately equal to  its argument for $|\bm\nabla\psi|\ll a_0$, and to level off at a value $\mu_0$ when $|\bm\nabla\psi|\gg a_0$.  Then for strong $|\bm\nabla\psi|$  Poisson's equations gives $\psi=\Phi_N/\mu_0$ so that the true potential is $\Phi=(1+1/\mu_0)\Phi_N$.  This means gravity is Newtonian with effective gravitational constant $G_N=(1+1/\mu_0)G$.  For $|\bm\nabla\psi|\ll a_0$,  the RAQUAL equation tells us that $\bm\nabla\psi$ is much stronger than $\bm\nabla\Phi_N$.   Consequently, by Eq.~\reff{sum},  $\bm\nabla\Phi\approx \bm\nabla\psi$ so that in the weak field regime RAQUAL's gravitational field is close to AQUAL's.  Thus in  RAQUAL we have a relativistic theory which in the nonrelativistic limit inherits the good traits of MOND by way of AQUAL.

However, on the relativistic front RAQUAL hit two roadblocks.  Already the original paper~\cite{BM}  remarks that perturbations of $\psi$ from a static background for which  $|\bm\nabla\psi|\ll a_0$, e.g. $\psi$ waves travelling in the reaches of a galaxy, propagate superluminally, i.e., outside the light cone of $\tilde g_{\alpha\beta}$.  Nowadays one hears the view that superluminality by itself is not sufficient reason for causal problems~\cite{bruneton}.  This is by no means the standard view, and so the specter of acausality motivated disillusionment with RAQUAL~\cite{Can,B90}, and formulation of a two-scalar field theory, phase-coupled gravity 
(PCG)~\cite{Can,PCG} intended to prevent superluminality.  In the PCG Lagrangian the new scalar $\sigma$ contributes a quadratic kinetic part, just as does $\psi$; there is also a simple coupling between $\sigma$ and the kinetic term for $\psi$ which gives the theory its MOND character.    Although  PCG generates an intolerable drift of the Kepler constant $a^3/P^2$ in the solar system, and marginally contradicts the observed precession of Mercury's perihelion~\cite{Can}, the theory has merits for both galaxy dynamics and cosmology~\cite{SandersPCG1,SandersPCG2}.    However it may be, RAQUAL and PCG were both finally disqualified as consistent representations of gravity by their inability to cope with observations of gravitational lenses. 

Discovered in 1979, gravitational lenses have been intensively studied ever since.  In strong gravitational lensing a cluster of galaxies (but sometimes a single galaxy) forms a few images of a single quasar in the background by  bending the incoming light rays with its gravitational field.   This provided a novel tool for determination of masses of large astronomical systems by an elementary use of GR.  By the late 1980's it was apparent that masses of clusters so determined were larger than the observed baryonic mass in them, and commensurate with the masses determined from motions of the galaxies comprising them.    The novelty was that the acceleration discrepancy was now put in evidence by a relativistic phenomenon.  Many pounced upon the finding as added support for DM's presence in clusters.

RAQUAL cannot deal with the new evidence; the reason is instructive.
 Maxwell's equations are known to be conformally invariant: replacement of the metric $g_{\alpha\beta}$ with which they are formulated by $f(x^\gamma)\,g_{\alpha\beta}$ does not change the form of the equations.    It follows that study of light propagation cannot distinguish between the metrics $g_{\alpha\beta}$ and $\tilde g_{\alpha\beta}$ of RAQUAL.  Accordingly, although Maxwell's equations are supposed to be written with $\tilde g_{\alpha\beta}$, the metric for studying gravitational lensing might just as well be $g_{\alpha\beta}$.   But this last obeys Einstein's equations whose sources are the baryonic matter's energy-momentum tensor and that of $\psi$.  However, for systems as distended as clusters or galaxies, this last is negligible.  The light bending, then, is predicted by RAQUAL to be nearly the same as predicted by GR \emph{without} DM.  Yet there is observational evidence that the inner parts of clusters  gravitationally lense light more strongly than would be expected from GR \emph{with no} DM.   The same problem recurs in PCG,  in which the two metrics are, again, conformally related in the same manner as in RAQUAL.

 It is thus quite clear that if a scalar field is to be the vehicle for MOND effects, then regardless of the form of its dynamics, the relation between the metrics $\tilde g_{\alpha\beta}$ and $g_{\alpha\beta}$ must be non-conformal.  I attempted ~\cite{B92} to construct such a coupling out of scalar field alone, to wit
 \begin{equation}
 \tilde g_{\alpha\beta}=e^{2\psi/c^2}(A\,g_{\alpha\beta}+B\, \psi_{,\alpha}\, \psi_{,\beta}),
 \end{equation}
 with $A$ and $B$ functions of the invariant $g^{\alpha\beta}\phi_{,\alpha}\,\phi_{, \beta}$.  However, Sanders and I found that avoidance of propagation of gravitational waves which are superluminal with respect to the matter metric $\tilde g_{\alpha\beta}$  requires that $B<0$, while $A>0$ is required so that both metrics have Lorentzian signatures~\cite{BS}.  We then calculated that for equal sources of Einstein's equations, the light bending in the proposed modification of RAQUAL is  \emph{weaker} than that in GR.  As mentioned,  in systems of the scale of galaxies and clusters of galaxies, the scalar field provides very little energy or momentum density as compared to the matter.  Thus in regard to light ray bending, the new theory performs worse than RAQUAL.
 
 Actually the above conclusion could be avoided if $\psi_{,\alpha}$ is a \emph{timelike} 4-vector.   However, we would expect that deep inside or near a galaxy, or a massive cluster of galaxies which is certainly virialized, the scalar field, whose principal source is the matter therein, should be quasistatic.  Then $\psi_{,\alpha}$ is expected to be spacelike, i.e. to give lensing weaker than needed.   The problem is thus still present. 
 
To overcome it Sanders~\cite{S97} proposed to replace $\psi_{,\alpha}$ in its above role by a nondynamical vector ${\cal U_\alpha}$ which in an isotropic cosmological model points precisely in the time direction, and approximately so in the presence of mass concentrations.  He further related the metrics  by
  \begin{equation}
 \tilde g_{\alpha\beta}=e^{-2\phi/c^2}\,  g_{\alpha\beta}- (e^{2\phi/c^2}-e^{-2\phi/c^2})\, {\cal U}_\alpha\,{\cal U}_\beta.
 \label{gtilde}
 \end{equation}
Sanders further supposed the action for the scalar field $\phi$, which supersedes $\psi$,  to be of RAQUAL form, or more generally with the function $F$ depending also on the scalar ${\cal U}^\alpha\,\phi_{,\alpha}$.   The timelike nature of the vector is conducive to enhancement of the lensing, and such a theory, named stratified theory, can cope with the observations of gravitational lensing.  However, it was clear to Sanders that this is just a toy theory since a globally timelike vector field, such as ${\cal U}_\alpha$,  determines a universal preferred frame of reference, an aether.   Additionally, since the vector field is supposed constant, the theory is not covariant. 

\section*{7. T$e$V$e$S and other relativistic MOND theories}

The mentioned problems are avoided by making the vector field a dynamical one,  as first done in T$e$V$e\,$S, my tensor-vector-scalar theory of gravity~\cite{B04,JHopk,BS2}.  T$e$V$e\,$S retains the relation~\reff{gtilde} between a gravitational metric $g_{\alpha\beta}$ and a physical (or observable) metric $\tilde g_{\alpha\beta}$.  The matter Lagrangian is built exclusively with $\tilde g_{\alpha\beta}$; this guarantees that the weak equivalence principle will be obeyed precisely.  The conventional  GR Einstein-Hilbert Lagrangian is used to give dynamics to $g_{\alpha\beta}$, which then trivially induce dynamics for $\tilde g_{\alpha\beta}$.    The scalar field $\phi$ is provided with a RAQUAL type Lagrangian.  For convenience this is couched in terms of a  quadratic form in the 4-gradient of $\phi$ in interaction with a second scalar field, $\sigma$, which also occurs in a potential-like term.  (However, no derivatives of $\sigma$ enter, so $\sigma$ is not dynamical, and if eliminated, the explicit aquadratic Lagrangian for $\phi$ shows up.) The mentioned quadratic form is not the usual one, but rather $(g^{\alpha\beta}-{\cal U}^\alpha\,{\cal U}^\beta) \phi_{,\alpha}\,\phi_{,\beta}$; it is introduced to  forestall the superluminal propagation that afflicts RAQUAL.

The kinetic part of the ${\cal U}_\alpha$ Lagrangian is quadratic in its first derivatives, and exactly analogous to that for a gauge field.  This is not the only quadratic form possible, but it is the one in which the Einstein equations for $g_{\alpha\beta}$ will not have second derivatives of ${\cal U}_\alpha$ in their sources. As mentioned, it is imperative that ${\cal U}_\alpha$ be a timelike vector.  In T$e$V$e\,$S this is accomplished by introducing, alongside the kinetic term, a Lagrange multiplier term of the form $\lambda({\cal U}_\alpha {\cal U}^\alpha+1)$ which forces the norm of ${\cal U}_\alpha$ to be negative (and ${\cal U}_\alpha$ to be of unit length as a bonus).

Whereas RAQUAL is a one parameter ($a_0$) theory with one free function ($\mu$), 
T$e$V$e\,$S has one free function, $F(\sigma^2)$ [distinct from the $F$ in Eq.~\reff{lagrangian}] that determines the \emph{shape} of the potential-like term, and three parameters in addition.  One of these is a scale of length $\ell$ in the $\phi$ Lagrangian that sets the strength of the potential, while a dimensionless one, $k$, determines the scale of its argument.  The third parameter, $K$, also dimensionless,  sets the strength of the ${\cal U}_\alpha$ Lagrangian.  One can form the acceleration scale
\begin{equation}
a_0={\sqrt{3k}c^2\over 4\pi\ell},
\label{a0}
\end{equation}
which parallels the role of Milgrom's constant in MOND and AQUAL.    It is found that the limit of T$e$V$e\,$S with $k\to 0$, $K\propto k$ and $\ell\propto k^{-3/2}$  amounts to GR.   In fact $a_0\to 0$ in this limit, so that all accelerations are to be regarded as strong, and gravitation as conventional.

How about nonrelativistic motion?  Let $\phi_c$ be the coeval value of $\phi$ in the cosmological model in which the nonrelativistic system is embedded.  This will be the asymptotic boundary value of $\phi$ for any local solution of the $\phi$ equation.   From a linearized version of Einstein's equations, and use of the relation~\reff{gtilde}, it follows~\cite{B04} that the effective nonrelativistic gravitational potential is
\begin{eqnarray}
\Phi&=&\Xi\Phi_N+\phi,
\label{sum2}
\\
\Xi &\equiv& (1-K/2)^{-1}\,e^{-2\phi_c/c^2}
\end{eqnarray}
(this corrects a sign error  in Refs.~\onlinecite{B04} and \onlinecite{JHopk}).
Here $\Phi_N$ comes from the usual Poisson equation while $\phi$ is found from the scalar equation; in both the source is the baryonic mass density $\rho$.  Thus far all studies have assumed that $0<K\ll 1$ and $|\phi_c|\ll c^2$, in which case  Eq.~\reff{sum2} essentially coincides with Eq.~\reff{sum} in RAQUAL.   For this same reason any time variation of $\Xi$~\cite{FamGen} should be negligible.  This is of particular significance for the above derived value of $a_0$ which, strictly speaking, includes a factor $\Xi$ [for example Eq.~(6.7) of Ref.~\onlinecite{JHopk}].
 
For a static situation the equation for $\phi$ takes a form like Eq.~\reff{AQUAL},  with $\mu$ (essentially $\sigma^2$)  expressed in terms of $|\bm\nabla\phi|$ through T$e$V$e\,$S's free function $F(\sigma^2)$.  By a suitable choice of   $F$ one can reproduce any $\mu$ function that would be relevant in AQUAL, with the scale thereon agreeing with $a_0$ in Eq.~\reff{a0}.  By analogy with our discussion of RAQUAL we see that T$e$V$e\,$S can have MOND phenomenology and also a Newtonian limit.  The transition between Newtonian and MOND regimes takes place as $|\bm\nabla\phi|$ sweeps through the value $a_0$, near the point where $|\bm\nabla\Phi|\approx a_0$.  

Actually for systems strongly departing from sphericity, the situation is not as clear, but it seems that at the nonrelativistic level T$e$V$e\,$S implements the MOND paradigm overall.  When it comes to details, my original choice for $F$~\cite{B04}  can be criticized:  while it  leads to flat outer RCs, it does not give a satisfactory account of the transition part of the RCs of several well measured galaxies, including our own~\cite{Binney,SandersSS}.  By contrast, the MOND formula with the ``simple'' variant of Migrom's $\mu$ (FIG.~\ref{figure:fig3}) leads to very good fits for many RC's~\cite{FamGen}.  Forms of the T$e$V$e\,$S  function $F$  which would do a correspondingly good job have been proposed by Zhao and Famaey~\cite{Famaey}, Sanders~\cite{SandersSS} and Famaey, et al.~\cite{FamGen}.

Sanders has also proposed a variation on T$e$V$e\,$S involving a second scalar $\sigma$ with PCG-like dynamics instead of RAQUAL ones ~\cite{BSTV}.  This bi-scalar tensor vector theory (BSTV) has three free functions and a free parameter.  The theory is not especially needed to obviate superluminal propagation since  T$e$V$e\,$S seems to do well on this~\cite{B04}.  However, BSTV, like PCG,  is a more appropriate frame for generating cosmological evolution of $a_0$.  Since numerically $a_0\sim cH_0$, it is often argued that $a_0$ must be determined by cosmology, and should thus vary on a Hubble timescale~\cite{M1,SMG}.  As mentioned, $a_0$ in T$e$V$e\,$S is essentially set by the parameters $\ell$ and $k$; it should be nearly constant in the expanding universe.  It is otherwise in BSTV.   Discrimination between constant and evolving $a_0$ may be possible with good RCs of disk galaxies at redshifts $z= 2$---5.  Such curves are just now coming into range.  In fact, the data in Ref.~\onlinecite{Genzel} put the $z=2.38$ galaxy BzK-15504  right on the  MOND derived Tully-Fisher law with the standard value for $a_0$~\cite{Mp}.  Thus the meagre data available today are consistent with no $a_0$ evolutions on the Hubble timescale.

Recently Zlosnik, Ferreira and Starkman~\cite{ZFS1} have clarified the relation between T$e$V$e\,$S  and the so called Einstein-Aether gravity theories in which a timelike unit vector field (but no scalar) plays a role alongside the metric~\cite{aether}.  They do this by re-expressing T$e$V$e\,$S  solely in terms of the observable metric $\tilde g_{\alpha\beta}$ and the vector field.  The $\phi$ is eliminated with the help of the constraint imposed through the Lagrange multiplier term.  There emerges an Einstein-Aether like theory in which the metric $\tilde g_{\alpha\beta}$ also satisfies Einstein-like equations.  However, in contrast to orthodox Einstein-Aether theories, the vector kinetic action in the theory in question is a generic quadratic form in the vector's derivatives contracted into polynomials of the vector's components.   This form of T$e$V$e\,$S is more complex than the original one, but there are some circumstances in which it  may be more convenient in ferreting out consequences of the theory.  

In  T$e$V$e\,$S in either form the vector is normalized with respect to $g_{\alpha\beta}$, not $\tilde g_{\alpha\beta}$.  Zlosnik, et al. have also proposed a variant tensor-vector theory with a timelike vector normalized with respect to $\tilde g_{\alpha\beta}$, and which also has MOND like behavior~\cite{ZFS2}.   This is constructed by taking the vector's action as a function ${\cal F}$ of ${\cal K}$, the  quadratic form in the derivatives of the vector field of orthodox Einstein-Aether theories.   The Zlosnik, et al. theory has four parameters:  a length scale and three dimensionless parameters.  The form of the free function ${\cal F}({\cal K})$ can be deduced approximately from the requirement that MOND arise in the nonrelativistic quasistatic limit, and from the stipulation that a cosmology built on this theory shall have an early inflationary period and an accelerated expansion at late times.

In summary, the relativistic implementations of MOND involve either one or three free functions; those with one free function have either three or four free parameters.

\section*{8. T$e$V$e\,$S and gravitational lensing}

As stressed earlier, the particular structure of  T$e$V$e\,$S  reflects the desire to encompass in the MOND paradigm the observation that what passes, from a dynamical point of view, for DM in galaxies and clusters of galaxies  also lenses light to a commensurate degree.   How does T$e$V$e\,$S  measure up to the task?  Now the measured gravitational lensing by galaxies and clusters of galaxies takes place over cosmological distances.  The  light rays in T$e$V$e\,$S are null geodesics of the observable metric $\tilde g_{\alpha\beta}$; to compute $\tilde g_{\alpha\beta}$  one needs $U_\alpha, \phi$ and $g_{\alpha\beta}$.   Isotropic cosmological models in T$e$V$e\,$S  closely resemble  the corresponding cosmological models of GR; they sport a $U_\alpha$ pointed precisely in the time direction and feature very slow change of $\phi$~\cite{B04}. A consequence is that $\tilde g_{\alpha\beta}$ in a T$e$V$e\,$S isotropic cosmological model differs little from the metric in the corresponding GR cosmology.  Thus the cosmological facet of gravitational  lensing is very much like in GR.  

The second facet is the local bending of light rays in a mass' vicinity.   As in isotropic cosmological models, so in static situations, the solution of the vector's equation has $U_\alpha$ pointed precisely in the time direction.  To compute the bending in linearized theory, the approximation commonly used in the business, one also needs the scalar field $\phi$ and the metric $g_{\alpha\beta}$, both to first order in $\Phi_N$.  The final result is that the line element takes the form
\begin{equation}
d\tilde s^2 = -(1+2\Phi/c^2)dt^2+(1-2\Phi/c^2)(dx^2+dy^2+dz^2)
\label{le}
\end{equation}
with $\Phi$ given by Eq.~\reff{sum2}.  Note that the same potential $\Phi$ appears in both terms of this isotropic form of the line element.  Hence light ray bending, which leans on both to equal degree, measures the same gravitational potential as do dynamics which are sensitive only to the temporal part of the line element.  This mirrors the situation in GR whose line element is obtained by sending $\Phi\mapsto \Phi_N$ in Eq.~\reff{le}. Thus half of the problem that plagues RAQUAL and similar theories is overcome: the acceleration discrepancy makes itself felt equally through the gravitational lensing as through the dynamics. 

The second half of the problem revolves about the mass distribution that generates $\Phi$.  In GR $\Phi$ is all there is, and its Laplacian, as determined from the lensing observations or from the dynamics,  will give the \emph{total} (baryonic plus dark) mass distribution directly.  Because DM is not seen directly, this prediction can only be judged by the plausibility of the derived distribution of DM.  By contrast in   T$e$V$e\,$S the measured $\Phi$ is to be decomposed into two parts in the manner of Eq.~\reff{sum2}, with each generated by the same \emph{baryonic} mass density $\rho$, one part through Poisson's equation, and the second through a highly nonlinear AQUAL type equation.   Evidently, in a T$e$V$e\,$S model of a gravitational lense, the baryonic matter will be distributed differently from the total matter in a GR model of the same lense.  And when the lensing system is not spherically symmetric, the centers of the two distributions may be offset.

Chiu, Ko and Tian~\cite{Chiu} have explored light ray bending by a pointlike mass $M$ in T$e$V$e\,$S.   They note that the deflection angle in the  deep MOND regime [impact parameter $b\gg b_0\equiv  (k/4\pi)(GM/a_0)^{1/2}$] approaches a constant, as might have been expected from naive arguments, but is less  predictable in the intermediate regime $b\sim b_0$.  Thus calculations of lensing based on a mixture of MOND and GR motives~\cite{MT,Clowe1} can easily mislead.  Chiu et al. work out the  lens equation in T$e$V$e\,$S, which controls the amplifications of the various images in the strong lensing of a distant source, and remark that for two images the difference in amplifications is no longer unity as in GR, and may depend on the masses.  With a photometric survey like the Sloan Digital Survey  it may be possible to check for this effect.  Finally, these authors work out the gravitational time delay in T$e$V$e\,$S, which could be of use in interpreting differential time delays in doubly imaged variable quasars.

More phenomenologically oriented,  Zhao, Bacon, Taylor and Horne~\cite{Zhao} compare T$e$V$e\,$S  predictions with a large sample of galaxy strong lenses which each produce two images of a quasar.  They  model the galaxy baryonic mass distribution either by point masses or by the popular Hernquist profile.  Galaxy masses are estimated by comparing observations both with predicted image positions and with predicted amplifications ratios; the two methods are found to give consistent results, themselves well correlated with the luminosities of the galaxies.  The corresponding mass-to-light ratios are found to be in the normal range for stellar populations, with some exceptions.

How frequently should strong lensing occur?  This question is taken up in the context of T$e$V$e\,$S by Chen and Zhao~\cite{Chen}.   Again modeling the mostly elliptical galaxies with Hernquist profiles, they compute the probability of two images occurring as a function of their separation.  Their prediction falls somewhat beneath the frequency observed in the lensing surveys, though they consider this still acceptable.  The predictions are sensitive to the assumed mass profile, as well as to the assumed shape of the $\mu$ function, which Chen and Zhao assume to switch brusquely from linear in the argument to unity.

\begin{figure}[ht]
\centering
\includegraphics[width=8cm]{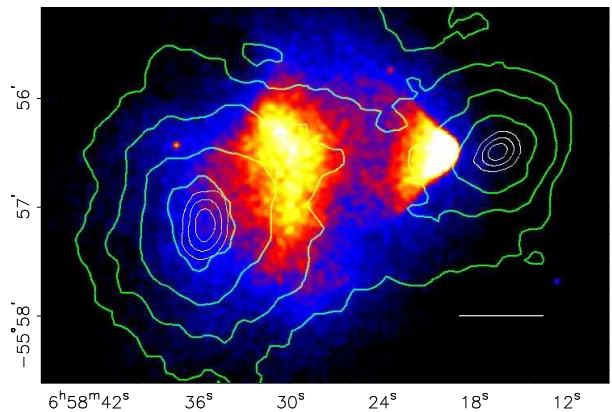}
  \caption{The colliding clusters 1E0657-56.  The bullet cluster (right) rammed through the cluster on the left.  Hot gas stripped off both clusters is colored red-yellow.  Green and white curves are level surfaces of gravitational lensing convergence; the two peaks of this do not coincide with those of the gas which makes up most of the visible mass, but are skewed in the direction of the galaxy concentrations. The white bar corresponds to 200 kpc.  Figure reproduced from Ref.~\onlinecite{Clowe2} by permission of the American Astronomical Society.}
 \label{figure:fig6}
 \end{figure}

Gravitational lensing by the colliding galaxy clusters 1E0657-56 has been claimed to give theory independent proof of DM dominance at large scales~\cite{Clowe1}.  In this system (FIG.~\ref{figure:fig6}) a smaller cluster, the ``bullet'', has crashed through a larger one and the intracluster gas of both has been stripped by the collision, the bullet's  gas trailing behind its galaxy component.  Weak lensing  (distorted but unsplit images) mapping  shows the lensing mass to be concentrated in the two regions containing the galaxies, rather than in the two clouds of stripped gas which contains the lion's share of baryonic mass~\cite{Clowe1,Clowe2}.  Collisionless DM would indeed move together with the galaxies. Hence the inference that much DM continues to accompany the bullet.   Angus, Famaey and Zhao~\cite{Angus1} (see also Ref.~\onlinecite{Famaey}) point out that in the very asymmetric system 1E0657-56, MOND, AQUAL and  T$e$V$e\,$S all predict substantially different gravitational field distributions (compare Eqs.~\reff{MOND} and \reff{integral}), a situation which confuses Clowe et al.'s ``theory independent'' inference.  

Whereas Angus, Famaey and Zhao consider it possible to explain the lensing with a reasonable purely baryonic matter distribution, a later paper by Angus, Shan, et al.~\cite{Angus2} concludes that dark matter is needed after all.  This is hardly surprising; as we saw in Sec.~3, pure MOND does not fully account
for the acceleration discrepancy in the dynamics of quiescent galaxy clusters~\cite{SMG}.  But  DM models of the bullet clusters within GR are not without their problems.  Farrar and Rosen~\cite{Farrar} note that the relative velocity of the clusters is too high as compared to those seen in DM simulations of structure formation.  To remove the contradiction they propose that a non-gravitational attraction of a new sort acts only between clumps of DM.  But is assuming existence of DM together with a new interaction specific to it more parsimonious than a modification of standard gravity such as MOND?

\section*{9.  T$e$V$e\,$S  and cosmology}
 
With my original choice of $F$, T$e$V$e\,$S cosmological models with baryonic matter content alone can be very similar to the corresponding GR models by virtue of the scalar $\phi$'s energy density remaining relatively small~\cite{B04,Chiu,Zhao}.   In particular Chiu et al.~\cite{Chiu}  note that these T$e$V$e\,$S models give a reasonable relation between redshift and angular distances, and are thus as effective as GR models in providing the scaffolding for the analysis of cosmologically distant gravitational lenses.  How would changing $F$ affect the evolution of a cosmological model?  This is studied exhaustively by Bourliot et al.~\cite{Bourliot} following an earlier exploration by Skordis et al.~\cite{Ferreira}.  Bourliot et al.  display a large set of variants of my $F$ for which, while the scalar energy density may not be negligible, it  tracks or mimicks the behavior of another energy component of the model, e.g. the radiation's during radiation dominance. They also characterize shapes of $F$ which can lead to future singularities in cosmology, and which presumably should be  avoided. 

MOND critics have always held up the complicated power spectrum of cosmological perturbations (background radiation or baryons) as a proof that DM is needed on the cosmic level; after all the spectrum is said to  be well fit by the ``concordance'' DM model of the universe.   This argument took for granted that MOND could never measure up to the test.  With T$e$V$e\,$S on the scene one can face the question technically.   
 
 In a massive work Skordis has provided the full covariant formalism for evolution of cosmological perturbations in T$e$V$e\,$S~\cite{Skordis}.  And using this Skordis et al.~\cite{Ferreira} have shown that, without invoking dark matter, T$e$V$e\,$S can be made consistent with the observed spectrum of the spatial distribution of galaxies and of the cosmic microwave radiation if one allows for contributions by massive neutrinos (in the still allowed mass range) and the cosmological constant.  In this approach the role of dark matter in standard cosmology is taken over by a feedback mechanism involving the scalar field perturbations.  Dodelson and Liguori~\cite{Dodelson} have independently calculated perturbation growth, and stressed that it is rather the vector field in T$e$V$e\,$S which is responsible for growth of large scale structure without needing DM for this~\cite{Dodelson}.  It thus seems there is potential in T$e$V$e\,$S to do away with cosmic DM, as well as with DM in galaxies.
 
 Much attention is commanded today by the mystery of the ``dark energy'', the agent responsible for the observed acceleration of the Hubble expansion.  Such agent is obligatory in the context of GR cosmological models.  Quite naturally many have wondered whether T$e$V$e\,$S could obviate this need.   Diaz-Rivera, Samushia, and Ratra~\cite{Diaz} have found exact deSitter solutions of T$e$V$e\,$S cosmology which can represent either early time inflation epochs or the late time acceleration era. In another T$e$V$e\,$S study, Hao and Akhoury~\cite{Hao} have concluded that with a suitable choice of the T$e$V$e\,$S function $F$, the scalar field can play the role of dark energy.   And Zhao~\cite{Zhao2} maintains that with Zhao and Famaey's choice of $F$~\cite{Famaey}, cosmological models can be had that evolve at early times like those of standard cold dark matter cosmology, and display late time acceleration with the correct present  Hubble scale, all this without assuming DM or dark energy.  For the related Einstein-Aether theory, Zlosnik et al.~\cite{ZFS2} remark that with suitable choice of  their theory's ${\cal F}$, the vector field can both drive early inflation as well as double for dark energy at late times.  
\section*{10.  Assessment of relativistic MOND}

It may be unnecessary to go far from Earth to tell T$e$V$e\,$S, BSTV and similar theories apart from GR; the solar system can be turned into a sieve for the correct modified gravity.  Between any two bodies in the solar system there is an extremum point, strictly a saddle point, of some relevant field.  In AQUAL this would be the $\Phi$ potential and in T$e$V$e\,$S the $\phi$ field.  Near such a point the gradient of the said field is small, and departures from Newtonian gravity are significant ($\mu$ in AQUAL falls short of unity, and the derivative of $F$ in T$e$V$e\,$S is driven away from the pole which signals Newtonian behavior).  A detailed study by Magueijo and me~\cite{Mag} with T$e$V$e\,$S shows that the anomalous regions, e.g. between Sun and Jupiter and Earth and Moon, are small, but gives some hope that with the fine guidance of space probes like ESA's LISA-Pathfinder (scheduled for liftoff in October 2009), it may be possible to diagnose MOND-like effects.

Well away from the saddle points, but still in the region occupied by the planets, T$e$V$e\,$S predicts small departures of  $\phi$ from $1/r$ form, i.e. small departures from Newtonian behavior as encoded in $\Phi$~\cite{B04}.  Similar predictions issue from BSTV~\cite{SandersSS}.    These are constrained by the observed constancy of Kepler's constant $a^3/P^2$ out to the orbits of the major planets, and by the absence of significant departures from GR's predictions for the precessions of the perihelia  of Mercury and the asteroid Icarus.  According to Sanders, T$e$V$e\,$S can satisfy these constraints with choices of the function the $F$ proposed by Refs.~\onlinecite{SandersSS} and~\onlinecite{Famaey}, which, as mentioned in Sec.~7,  also lead to correct galaxy RCs.  BSTV can do just as well when its extra parameter, $\epsilon$, is chosen properly~\cite{SandersSS}.

Both T$e$V$e\,$S and BSTV with the above choices predict that beyond 100 astronomical units (AU) from the sun, there is a fairly $r$-independent attractive component of the sun's gravitational field of strength $\sim 0.3 a_0$.  The Pioneer anomaly, as measured at distances beyond 20  AU, is also $r$-independent but stronger.  Is this evidence for MOND?  Unfortunately, the interpretation of the Pioneer anomaly as an almost constant sunward acceleration seems to clash with limits on the variation of Kepler's constant set with spacecraft at the distance of Uranus and Neptune~\cite{Anderson95,SandersSS} and with the latest ephemerides of the outer planets~\cite{Iorio}.  It thus seems prudent to suspend judgment until a prosaic origin for Pioneer anomaly, e.g. drag by unknown matter,  can be excluded with high probability. 

The above focuses on the nonrelativistic limit.  How do relativistic MOND theories fare with regard to the lowest order relativistic effects in the solar system?  As for any metric theory of gravity, these effects can be calculated from the post-Newtonian (PPN) parameters of the theory of choice.   For any metric theory it is possible to parametrize the first and second order departures of the metric from Minkowski form in terms of a set of ten intuitive looking potentials~\cite{Will}.  The Newtonian potential $\Phi_N$  figures in the list, and occurs in the corrections both linearly as well as squared; two others are
\begin{equation}
\Phi_2=\int{\rho({\bm r'})\Phi_N ({\bm r'})\over |{\bm r}-{\bm r}'|}d^3r',\quad V_i=\int{\rho({\bm r'})v_i({\bm r'})\over |{\bm r}-{\bm r}'|}d^3r',
\end{equation}
where ${\bm v}$ denotes the fluid velocity in the system.
The post-Newtonian (PPN) parameters are the dimensionless coefficients multiplying the diverse potentials in the correction.  Some examples will clarify this.  

The correction to the space-space part of the metric, $g_{ij}$,  is $2\gamma\Phi_N\delta_{ij}$; it defines the PPN parameter $\gamma$.  The correction to the temporal metric component, $g_{tt}$, starts with the terms $2\Phi_N-2\beta\Phi_N{}^2$, which define the PPN parameter $\beta$, and also includes a $\Phi_2$ term whose coefficient brings in a further PPN parameter, $\zeta_1$.  Each time-space metric component $g_{ti}$  gets corrections proportional to the $V_i$ and $w_i \Phi_N$ (${\bm w}$ denotes the velocity of the chosen coordinate system with respect to the cosmological matter comoving frame).  With these last terms come two additional PPN parameters, $\alpha_1$ and $\alpha_2$,  which are called preferred frame parameters.  A third one, $\alpha_3$, is associated with a correction of the form ${\bm w}^2\Phi_N$ to $g_{tt}$.  There are four more PPN parameters for a total of ten.

The parameters $\beta$ and $\gamma$ are both unity  for GR and for some of it competitors. They have also been computed to be unity in T$e$V$e\,$S~\cite{B04,Giannios}, which thus fares as well as GR in reference to gravitational light bending, perihelia precessions of the planets, and the radar time delay.   Sanders~\cite{SandersSS} has provided a simple argument that $\beta$ and $\gamma$ are always unity in a class of theories (including BSTV) which start from the disformal relation~\reff{gtilde}.   

Next in order of relevance are  the preferred frame  parameters $\alpha_1, \alpha_2,$ and $\alpha_3$.  These vanish in GR; after all, GR has no preferred frames.  In T$e$V$e\,$S or BSTV at least one of the $\alpha$s should be nonvanishing because the vector ${\cal U_\alpha}$ establishes a locally preferred frame, with the vector pointing out the time direction in that frame. Sanders argues heuristically that  the $\alpha$s should be strongly suppressed in both theories~\cite{SandersSS}.   But explicitly computing the $\alpha$'s for relativistic MOND theories should be a high priority because they are subject to tight experimental bounds.

The odyssey in search of a relativistic embodiment of the MOND paradigm has led, not  to one relativistic MOND, but to many.  T$e$V$e\,$S by itself exemplifies a family of theories. First there is the freedom in the choice of the function $F$, which is still only modestly constrained~\cite{Binney,Famaey,SandersSS}.   Next, the coefficient of the second term in the factor $(g^{\alpha\beta}-{\cal U}^\alpha {\cal U}^\beta)\phi_{,\alpha}\phi_{\,\beta}$ in the scalar's Lagrangian can be changed to any other negative number.  In T$e$V$e\,$S the Lagrangians for $\phi$ and ${\cal U}_\alpha$ are formulated upon the metric $ g_{\alpha\beta}$; a different but related theory emerges if one uses instead the background of $ \tilde g_{\alpha\beta}$. Additionally  one can replace the aquadratic Lagrangian for $\phi$ by a PCG-style Lagrangian for two scalars, as indeed done in Sanders' BSTV.  Finally, one can  altogether  dispense with the scalar fields and go the way of the Zlosnik-Ferreira-Starkman theory~\cite{ZFS2}.

While the lack of uniqueness of relativistic MOND is a nuisance, it does add needed flexibility to the search for a ``final'' fundamental theory with which to underpin the MOND paradigm.  It is already clear that some of the above mentioned theories may not be in full accord with the facts.  For example, the published T$e$V$e\,$S has an exact MOND limit at low accelerations, yet as mentioned in Sec.~3, MOND cannot handle the dynamics of clusters of galaxies without invoking additional unseen matter.  An interesting escape is suggested by Sanders~\cite{Aegean}. In BSTV cosmology, just as in the PCG one~\cite{SandersPCG2},  the $\sigma$ field can undergo oscillations which generate bosonic particles early on.  If the BSTV parameters are right, these bosons can be trapped by clusters (but not by galaxies), and can thus comprise the additional unseen matter.  The charm of this resolution is that this new cold dark matter emerges from the modified gravity theory itself,  and is not a separate invention.

Relativistic MOND as here described has developed from the ground up, rather than coming down from the sky: phenomenology, rather than pure theoretical ideas, has been the main driver.  Actually a large industry flourishes on the sidelines with imaginative ideas from first principles regarding the essence of MOND.  I have not touched here on these motley approaches because  they have given so little that is observationally viable.  Neither have I dwelt here on modified gravity theories  that are not MOND motivated or oriented.  But the time may be ripe for turning to a more deductive approach to MOND.  Now that we have some idea of what constitutes a viable relativistic MOND theory, it should be easier to single out theoretical frameworks which might yield a promising candidate for the fundamental MOND theory either as a limiting case, as an effective theory, after dimensional reduction, etc. 
 
 To give an example, let us recall that Brans-Dicke modified gravity plus Maxwellian electromagnetism in the real world can both be recovered by dimensional reduction of 5-D Einstein gravity theory.  T$e$V$e\,$S has the same number of degrees of freedom as that pair of theories; specifically, its vector field has three degrees of freedom on account of the normalization condition, and the electromagnetic vector potential also has three on account of gauge freedom. Might some variant of T$e$V$e\,$S arise from dimensional reduction of a pure gravity theory in 5-D?  If so, this would both ameliorate the common feeling that T$e$V$e\,$S is unduly complicated, and point the way to a lode for theories which might do better justice to the observations in the spirit of the MOND paradigm.
 
  \section*{ Acknowledgments}   I thank Mordehai Milgrom, Bob Sanders and Stacy McGaugh for many useful remarks on the original manuscript, and the last two as well as Douglas Clowe for providing figures.  Research on this subject has been supported by grant 694/04 from the Israel Science Foundation established by the Israel Academy of Sciences and Humanities.


\begin{quote}
\emph{Jacob Bekenstein} received his Ph. D. (1972) from Princeton University.  He worked at the Ben Gurion University in Israel, where he became full professor in 1978 and the Arnow Professor of Astrophysics in 1983.  In 1990 he moved to the Hebrew University of Jerusalem where he is the Polak Professor of Theoretical Physics.  A member of the Israel Academy of Sciences and Humanities, of the World Jewish Academy of Sciences, and of the International Astronomical Union, Bekenstein is a laureate of the Rothschild prize and of the Israel National Prize.  His scientific interests include gravitational theory, black hole physics, relativistic magnetohydrodynamics, galactic dynamics, and the physical aspects of information theory.
\end{quote}
\end{document}